\documentclass[epj,final]{svjour}

\usepackage{latexsym}
\usepackage{url}
\usepackage{amsfonts}
\usepackage{amsmath, amssymb}
\RequirePackage{graphicx}

\begin{document}
\title{Gravitational waves as waveguides}
\author{ A.A.Kocharyan\inst{1}, M.Samsonyan\inst{2}, V.G.Gurzadyan\inst{2,3}
}                     
%
%
\institute{School of Physics and Astronomy, Monash University, Clayton, Australia \and Center for Cosmology and Astrophysics, Alikhanian National Laboratory and Yerevan State University, Yerevan, Armenia \and 
SIA, Sapienza Universita di Roma, Rome, Italy}
\date{Received: date / Revised version: date}
%

\abstract{We show that gravitational waves can act as waveguides for electromagnetic radiation, that is if the latter is initially aligned with the gravitational waves, then the alignment will survive during the propagation. The analysis is performed using the Hamiltonian formalism and the Jacobi equation for null geodesics and conditions for certain cases of polarization of the waves are obtained. The effect of waveguiding by the gravitational waves can influence the interpretation of associated gravitational and electromagnetic wave events, since the latter cannot necessarily obey the inverse square decay law for intensity.} 
%
\maketitle

\section{Introduction}

The experimental discovery of the gravitational waves \cite{Ligo} (GW) opened an entire new window in the study of the Universe. GW data are used to address the key cosmological problems and tensions, constrain fundamental physical parameters, modified gravity theories, complementing the observational and experimental surveys, e.g. \cite{Ligo1,Mu,C4,Cap,Mas} and references therein. Particular importance have the associated events of both gravitational and electromagnetic (EM) wave pulses, as it was with GW170817 \cite{Troja}, providing information on the merging details, jets and cosmology \cite{Si}. In view of the possible observations in future of more counterparts of GW and EM sources, the consideration of effects of their associated propagation becomes important. Let us note, that waveguiding of GW by matter, lensing effects, has been considered earlier (\cite{Cap1,Cap2} and references therein), while we study the trapping and waveguiding of EM by GW themselves.

We will study the impact of the gravitational wave propagation on the aligned photons by means of the Hamiltonian formalism, null geodesic congruences, the deviation of null geodesics and the hyperbolicity of geodesic flows \cite{An,Arn,HE}. That formalism appears to be efficient in the studying the properties of the Cosmic Microwave Background (CMB), Cold Spot, photon propagation through cosmic voids \cite{GK1,GK2,spot,S2}, linked to the Hubble tension, e.g. \cite{GS7,GS8}. We show that, within certain conditions, the photons if initially aligned with the gravitational waves, will continue to propagate keeping their alignment, i.e. staying trapped within the gravitational waves. This is a principal conclusion, since it implies that gravitational waves can possess properties of waveguides, transmitting electromagnetic radiation not determined by the inverse square law decay of intensity.

\section{Deviation of null geodesics}

We start from the deviation of null geodesics defined via the Jacobi equation \cite{An,HE}
\begin{equation}
  \frac{d^2n^a}{dt^2}+R^a{}_{bcd}u^b u^d n^c=0,
\end{equation}
where $u$ and $n$ denote the velocity and the deviation of geodesics, respectively. This equation shows that the behavior of deviation vector $n^a$ depends explicitly on Riemannian tensor $R_{abcd}$. The association of the null geodesics and plane gravitational waves are known since the seminal study by Penrose \cite{Pen}.

Jacobi equation (1) can be represented as follows \cite{HE} (Eqs. (4.34)-(4.36) there, we keep the same notations)
\begin{align}
\dot{\theta}
&= -R_{ab}n^an^b+2\omega^2 -\sigma_1^2-\sigma_2^2-\tfrac12\theta^2,\\
\dot{\omega}
&=-\theta\ \omega,\\
\dot{\sigma_1}&=-\theta\ \sigma_1-C_{1010},\\
\dot{\sigma_2}&=-\theta\ \sigma_2-C_{1020},
\end{align}where
\begin{align}
\theta=(\det A)^{-1}\frac{d}{ds}(\det A),
\end{align}
\begin{align}
\hat{\omega}_{mn}=
\begin{pmatrix}
&0 &\omega\\
&-\omega &0
\end{pmatrix},\qquad
\hat{\sigma}_{mn}=
\begin{pmatrix}
&\sigma_1 &\sigma_2\\
&\sigma_2 &-\sigma_1
\end{pmatrix}
\end{align}
and $C_{abcd}$ is the Weyl tensor,  $\theta$ is the expansion scalar as the trace of the expansion tensor, $\omega_{mn}$ is the vorticity tensor, $\sigma_{mn}$ is the shear tensor, the matrix $A$ is defining the shape and orientation of the fluid element. Eq.(2) is known as Landau--Raychaudhuri equation for null geodesics \cite{HE}. 

We will consider the case with $\omega_{mn}=0$ (no vorticity, i.e. no centrifugal forces) to obtain the mutual behavior of $\theta$ and $\hat\sigma$ vs the affine parameter of the geodesics, and hence, constraints on the alignment of the photon bundle during the propagation.

We consider gravitational waves in empty space-time, hence we have (cf. \cite{MTW}, Chapter 35)
\begin{align}
    R_{ab}&=0,\\
    C_{abcd}&=R_{abcd}.
\end{align}

Denoting $a_1=R_{1010}$ and $a_2=R_{1020}$, we have

\begin{align}
\dot{\theta}
&= 2\omega^2 -\sigma_1^2-\sigma_2^2-\tfrac12\theta^2,\\
\dot{\omega}
&=-\theta\ \omega,\\
\dot{\sigma_1}&=-\theta\ \sigma_1-a_1,\\
\dot{\sigma_2}&=-\theta\ \sigma_2-a_2.
\end{align}
If $\theta=2\dot{\ell}/\ell$, then 
\begin{align}
	\dot\theta+\tfrac12\theta^2&=2\frac{\dot{\ell}}{\ell}
\end{align}
and we get a closed equation for $\ell$

\begin{align}
\ddot{\ell}&=\left[\omega^2-\tfrac12(\sigma_1^2+\sigma_2^2)\right]\ell,
\end{align}
where
\begin{align}
\omega&=\omega_0\frac{\ell_0^2}{\ell^2},\\
\sigma_i&=\left(\sigma_{i,0}-\int_0^\lambda a_i\ell^2ds\right)\frac{\ell^2_0}{\ell^2},
\end{align}
and
\begin{align}
\ddot{\ell}&=\left(\omega_0^2
-\tfrac12\left[\left(\sigma_{1,0}-\int_0^\lambda a_1\ell^2ds\right)^2 
+\left(\sigma_{2,0}-\int_0^\lambda a_2\ell^2ds\right)^2\right]\right)\frac{\ell_0^4}{\ell^3}.
\end{align}

If $\omega=0$, then
\begin{align}
\dot{\theta}
&= -\sigma_1^2-\sigma_2^2-\tfrac12\theta^2,\\
\dot{\sigma_i}
&=-\theta\ \sigma_i-a_i.
\end{align}

We can make the following changes
\begin{align}
\dot{\theta}
&= -\tfrac12\left(\theta+\sigma_1+\sigma_2\right)^2 +\theta(\sigma_1+\sigma_2)
-\tfrac12\left(\sigma_1-\sigma_2\right)^2,\\
\dot{\sigma_1}
&=-\theta\ \sigma_1-a_1,\\
\dot{\sigma_2}
&=-\theta\ \sigma_2-a_2.
\end{align}
Denoting $\sigma=\frac12(\sigma_1+\sigma_2)$, $\mu=\theta +2\sigma$,  and $\delta=\sigma_1 -\sigma_2$, we will get
\begin{align}
\dot{\mu}
&= -\tfrac12\mu^2 -\tfrac12\delta^2-(a_1+a_2),\\
\dot{\sigma}
&=-\theta\ \sigma-\tfrac12(a_1+a_2),\\
\dot{\delta}
&=-\theta\ \delta-(a_1-a_2).
\end{align}

Below, we will analyse this set of equations determining the deviation of photon trajectories for certain special cases. 

\section{Photon motions in special cases}

\subsection{Special case 1 ($\sigma_1=\sigma_2$)}
If $a_1=a_2=a$ ($R_{1010}=R_{1020}$), then our set of equations becomes

\begin{align}
\dot{\mu}
&= -\tfrac12\mu^2 -\tfrac12\delta^2-2a,\\
\dot{\sigma}
&=-(\mu-2\sigma)\ \sigma-a,\\
\dot{\delta}
&=-(\mu-2\sigma)\ \delta.
\end{align}

and the following solution
\begin{align}
    \theta(t)&=\sqrt{a}\left[\tanh(\sqrt{a} (t-t_1))-\tan (\sqrt{a}(t-t_0))\right]\label{theta},\\
    \sigma(t)&=-\tfrac{1}{2} \sqrt{a}\left[\tanh(\sqrt{a} (t-t_1))+\tan (\sqrt{a}(t-t_0))\right]\label{sigma},\\
    \sigma_1(t)&=\sigma_2(t)=\sigma(t),\\
    \delta(t)&=0.
\end{align}
\begin{figure}[h!]
	\caption{Photon trapping in gravitational waves indicated by the mutual behavior of $\theta$ and $\sigma$ during the propagation.  The scalar $\theta$ determines the rate of the change of volume element in time measured by a comoving observer, while the shear $\sigma$ determines the distortion of the shape of an initial ball.}
	\centering
	\vspace{5mm}
	\includegraphics[scale=0.5]{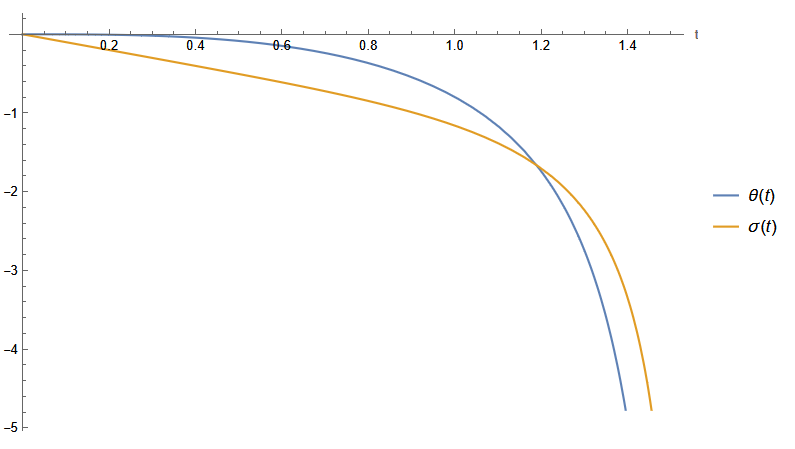}
\end{figure}
\subsection{Special case 2 ($\sigma_1/a_1=\sigma_2/a_2$)}

In this case we have
\begin{align}
\dot{\theta}
&= -\sigma_1^2-\sigma_2^2-\tfrac12\theta^2,\\
\dot{\sigma_1}
&=-\theta\ \sigma_1-a_1,\\
\dot{\sigma_2}
&=-\theta\ \sigma_2-a_2,
\end{align}
and
\begin{align}
\frac{\dot{\sigma_1}}{a_1}&=-\theta\ \frac{\sigma_1}{a_1}-1,\\
\frac{\dot{\sigma_2}}{a_2}&=-\theta\ \frac{\sigma_2}{a_2}-1,
\end{align}
hence
\begin{align}
\frac{\dot{\sigma_1}}{a_1}-\frac{\dot{\sigma_2}}{a_2}&=
-\theta\left(\frac{\sigma_1}{a_1}-\frac{\sigma_2}{a_2}\right).
\end{align}

If $\theta=2\dot{\ell}/\ell$, then 
\begin{align}
\ddot{\ell}&=-\tfrac12(\sigma_1^2+\sigma_2^2)\ell,\\
\left(\frac{\sigma_1}{a_1}\ell^2\right)^.&=-\ell^2,\\
\left(\frac{\sigma_2}{a_2}\ell^2\right)^.&=-\ell^2,\\
\left(\frac{\sigma_1}{a_1}-\frac{\sigma_2}{a_2}\right)\ell^2&=S.
\end{align}
For
$\nu_i=\sigma_i/a_i$ we have
\begin{align}
\ddot{\ell}&=-\tfrac12(a_1^2\nu_1^2+a_2^2\nu_2^2)\ell,\\
\left(\nu_1\ell^2\right)^.&=-\ell^2,\\
\left(\nu_2\ell^2\right)^.&=-\ell^2,\\
\left(\nu_1-\nu_2\right)\ell^2&=S.
\end{align}

If $S=0$, then $\nu_1=\nu_2=\nu$ and
\begin{align}
\ddot{\ell}&=-\tfrac12(a_1^2+a_2^2)\nu^2\ell,\\
\left(\nu\ell^2\right)^.&=-\ell^2,
\end{align}
or
\begin{align}
\dot{\theta}&=-\tfrac12\theta^2-(a_1^2+a_2^2)\nu^2,\\
\dot\nu&=-\theta\nu-1,
\end{align}
and
\begin{align}
    (\theta+2\alpha\nu)^.
    &=-\tfrac12\left((\theta+2\alpha\nu)^2+2(a_1^2+a_2^2-2\alpha^2)\nu^2\right)-2\alpha.
\end{align}

If $\alpha^2=\tfrac12(a_1^2+a_2^2)$, then
\begin{align}
    (\theta+2\alpha\nu)^.&=-\tfrac12(\theta+2\alpha\nu)^2-2\alpha,
\end{align}
which can be solved as in case 1.

If $a_1=a_2=a$, then $\alpha=a$ and $a\nu=\sigma$ (cf. case 1).

\subsection{Solution leading to the case $a_1=a_2$}

Equations that we have to solve can be written in a compact way
\begin{align}
&\dot{\sigma_i}+\theta\ \sigma_i
=-a_i\label{tosolve1},\\
&\dot{\theta}+\tfrac12\theta^2+\sigma_1^2+\sigma_2^2
= 0,\label{tosolve2}
\end{align}
where the index $i=1,2$

We can refer to $\sigma_1$ and $\sigma_2$ as a vector $\left( {\sigma_1\atop \sigma_2}\right)$, the rotation of which by an angle $\varphi$ will not change its length $\sigma_1^2+\sigma_2^2$ appearing in \eqref{tosolve2}. Let us denote the rotated vector by $\left({s_1\atop s_2}\right)$. Thus

\begin{align}
\left({s_1\atop s_2}\right)=
\begin{pmatrix}
&\cos\varphi &\sin\varphi\\
&-\sin\varphi &\cos\varphi
\end{pmatrix} \left({\sigma_1\atop \sigma_2}\right).
\end{align}
It follows from \eqref{tosolve1} that the vector $\left( {a_1 \atop a_2}\right)$ is also rotating
\begin{align}
\left({\tilde{a}_1\atop \tilde{a}_2}\right)=
\begin{pmatrix}
&\cos\varphi &\sin\varphi,\\
&-\sin\varphi &\cos\varphi
\end{pmatrix} \left({a_1\atop a_2}\right).
\end{align}
Let us choose the angle $\varphi$ such that $\tilde{a}_1=\tilde{a}_2$. Thus, we get
\begin{eqnarray}
a_1\cos\varphi +a_2\sin\varphi=-a_1\sin\varphi +a_2\cos\varphi
\end{eqnarray}
And hence
\begin{eqnarray}
\tan\varphi=\frac{a_2-a_1}{a_2+a_1}\label{angle}.
\end{eqnarray}
Denote $\tilde{a}_1=\tilde{a}_2=a$. In terms of $s_i$ and $a$, the equations \eqref{tosolve1} and \eqref{tosolve2} become
\begin{align}
&\dot{s_i}+\theta s_i
=-a\label{tosolve3},\\
&\dot{\theta}+\tfrac12\theta^2+s_1^2+s_2^2
= 0\label{tosolve4}.
\end{align}
When, in addition $s_1=s_2$, one goes back to $a_1=a_2=a$ case (Special case 1;\eqref{sigma}, \eqref{theta}).
\begin{eqnarray}
s(t)
&&=-\frac{\sqrt{a}}{2} \left[\tanh (\sqrt{a}(t-t_1))+\tan (\sqrt{a}(t-t_0))\right],      \label{st}\\
\theta(t)&&=\sqrt{a}\left[\tanh(\sqrt{a}(t-t_1))-\tan (\sqrt{a}(t-t_0))\right].     \label{thetat}
\end{eqnarray}
Then

\begin{eqnarray}
\sigma_1=(\cos\varphi-\sin\varphi)s,\\
\sigma_2=(\cos\varphi+\sin\varphi)s,
\end{eqnarray}
where $\varphi$ is given by \eqref{angle}.
Thus
\begin{eqnarray}
\sigma_1=\frac{a_1}{\alpha}\, s,\\
\sigma_2=\frac{a_2}{\alpha}\, s,\, 
\end{eqnarray}
where $s$ is given by \eqref{st}.

\section{Hamiltonian motion}

We will now use the Hamiltonian formalism to determine the behavior of the photon beams propagating within the gravitational waves defined by the linearized metric (weak-field approximation) \cite{MTW,W}
\begin{align}
	\mathbf{g} &= -\mathbf{d}t^2+\mathbf{d}z^2+\gamma_{ij}\mathbf{d}x^i\mathbf{d}x^j
    =-\mathbf{d}u\mathbf{d}v+(\delta_{ij}+h_{ij}(u))\mathbf{d}x^i\mathbf{d}x^j\qquad (h_{ij}(u)=A(u)e_{ij}),
\end{align}	
or
\begin{align}	
\mathbf{g}&=\begin{pmatrix}
    0 & -1/2 & 0 & 0\\
    -1/2 & 0 & 0 & 0\\
    0 & 0 & 1+h_{11}(u) & h_{12}(u)\\
    0 & 0 & h_{21}(u) & 1+h_{22}(u)
    \end{pmatrix}.
		\end{align}\\
		The Hamiltonian is defined as 
    \begin{align}
	\mathcal{H} &= -2p_up_v + \tfrac12(\delta^{ij}+A(u)e^{ij})p_ip_j =0,
	\end{align}
	where
	\begin{align}
    A(u)&=A_0\cos(\omega u).
\end{align}
Then
\begin{align}
    h_{11}&=-h_{22}=Re(A_+e^{-i\omega u}),\\
h_{21} &=h_{12}=Re(A_\times e^{-i\omega u}),
\end{align}
where $e^{ij}$ is the polarization tensor of the gravitational waves, and $u=t-z$ and $v=t+z$. 

The Hamiltonian equations then have the form, with the defined momentum $p_{i}$ 
\begin{align}
\begin{cases}
\frac{du}{d\lambda} = \frac{\partial\mathcal{H}}{\partial p_u}=-2p_v,\\
\frac{dv}{d\lambda} = \frac{\partial\mathcal{H}}{\partial p_v}=-2p_u,\\
\frac{dx^i}{d\lambda} = \frac{\partial\mathcal{H}}{\partial p_i}=\gamma^{ij}p_j,
\end{cases}
\end{align}
and
\begin{align}
\begin{cases}
\frac{dp_u}{d\lambda} = -\frac{\partial\mathcal{H}}{\partial u}=-\tfrac12A'(u)e^{ij}p_ip_j,\\
\frac{dp_v}{d\lambda} = -\frac{\partial\mathcal{H}}{\partial v}=0,\\
\frac{dp_i}{d\lambda} = -\frac{\partial\mathcal{H}}{\partial x^i}=0,
\end{cases}
\end{align}
thus $p_v=p_v(0)\ne0$, $p_i=p_i(0)$, and $du=-2p_v(0)d\lambda$, and
\begin{align}
\begin{cases}
\frac{dv}{du} = \frac{p_u}{p_v(0)},\\
\frac{dx^i}{du} = -\frac{\gamma^{ij}(u)p_j(0)}{2p_u(0)},\\
\frac{dp_u}{du} =\tfrac14A'(u)e^{ij}\frac{p_i(0)p_j(0)}{p_v(0),}
\end{cases}
\end{align}and
\begin{align}
\mathcal{H} &= -2p_up_v + \tfrac12(\delta^{ij}+A(u)e^{ij})p_ip_j =0,\\
 p_u&=\tfrac14(\delta^{ij}+A(u)e^{ij})\frac{p_i(0)p_j(0)}{p_v(0),}
\end{align}therefore
\begin{align}
\begin{cases}
u=u(0)-2p_v(0)\lambda,\\
\frac{dv}{du} = \frac{p_u}{p_v(0)},\\
\frac{dx^i}{du} = -\tfrac12(\delta^{ij}+A(u)e^{ij})\frac{p_j(0)}{p_v(0)},\\
p_u=\tfrac14(\delta^{ij}+A(u)e^{ij})\frac{p_i(0)p_j(0)}{p_v(0)},
\end{cases}
\end{align}
\begin{align}
\begin{cases}
u=u(0)-2p_v(0)\lambda,\\
v(u) =v(0)+\tfrac14(\delta^{ij}u+B(u)e^{ij})\frac{p_i(0)p_j(0)}{p_v(0)p_v(0)},\\
x^i(u) = x^i(0) -\tfrac12(\delta^{ij}u+B(u)e^{ij})\frac{p_j(0)}{p_v(0)},\\
p_u=\tfrac14(\delta^{ij}+A(u)e^{ij})\frac{p_i(0)p_j(0)}{p_v(0)},\\
B(u)=\int_0^uA(s)ds.
\end{cases}
\end{align}From
\begin{align}
p_u&=\tfrac12(p_t-p_z)\\
p_v&=\tfrac12(p_t+p_z)
\end{align}
for photons with $p_t(0)=p_z(0)$, $p_1(0)=p_2(0)=0$ we have $p_t(\lambda)=p_z(\lambda)$ for all $\lambda\ge0$. 

Therefore, if photons  are aligned with gravitational waves initially, then they will stay aligned during the propagation, trapped by the
gravitational waves.

It is interesting that, for small $p_i(0)$, $p_t\approx p_z$
\begin{align}
p_t-p_z&=(\delta^{ij}+A(u)e^{ij})\frac{p_i(0)p_j(0)}{p_v(0)}.
\end{align}
\begin{figure}[h!]
	\caption{Schematic view of the gravitational waves acting as waveguides for the aligned photons.}
	\centering
	\vspace{5mm}
	\includegraphics[scale=0.7]{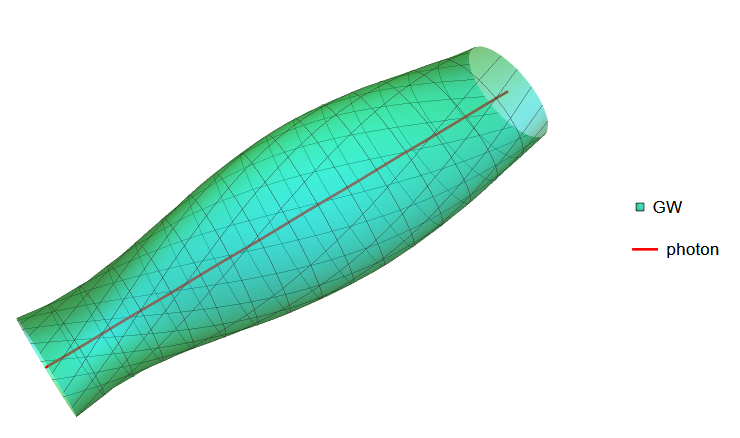}
\end{figure}

\section{Conclusions}

Using the Hamiltonian formalism and the Jacobi equation of the divergence of null geodesics (analog of Landau--Raychaudhuri equation), we analysed the propagation of
electromagnetic radiation in the metric of the gravitational waves. Particular cases of parameters of both waves are considered, and the conditions are obtained when the photons aligned with the gravitational wave initially, will continue to remain aligned, i.e. being trapped by the gravitational wave metric.
   
If the gravitational waves act as waveguides to transmit the electromagnetic waves, then the intensity of the latter will not decay by distance according to the inverse square law. This effect can be important especially at the
interpretation of the optical or X-ray counterparts associated to the gravitational wave signals  \cite{Troja}. Namely, then the detected intensity of electromagnetic radiation can indicate a lower integral power of its pulse, as compared to the power evaluated by the inverse square law. 

The considered gravitational waves' waveguiding effect can be particularly relevant at the accurate determination of the cosmological distance scale and the Hubble tension, as the detection of the first electromagnetic counterpart of the binary neutron star GW170817 already enabled such direct measurement of the Hubble constant \cite{Ligo2,Ligo3}. Further detection of such counterparts will provide an independent ruler for the cosmic ladder, complementing those of SN Ia, Baryonic Acoustic Oscillation (BAO), light element abundance measurements from Big Bang Nucleosynthesis (BBN) for low and high redshift samples \cite{BAO}.

\section{Data Availability Statement} 
Data sharing not applicable to this article as no datasets were generated or analysed during the current study.


\begin{thebibliography}{99}

\bibitem{Ligo} B. P. Abbott, et al. (LIGO Scientific Collaboration, Virgo Collaboration), Phys. Rev. Lett. 116, 061102 (2016)
\bibitem{Ligo1} R. Abbott, et al. (LIGO Scientific Collaboration, Virgo Collaboration, KAGRA Collaboration), Phys. Rev. D 104, 022004 (2021)
\bibitem{Mu} T. W. Murphy, Rep. Prog. Phys. 76, 076901 (2013)
\bibitem{C4} I. Ciufolini, et al, Eur. Phys. J. C,  76, 120 (2016)
\bibitem{Cap} S. Capozziello, M. Capriolo, Class. Quantum Grav. 38 175008 (2021)
\bibitem{Mas} S. Mastrogiovanni et al, Ann. Phys. (Berlin), 2200180 (2022)
\bibitem{Troja} E. Troja et al, Nature, 551, 71 (2017)
\bibitem{Si} A. Singhal et al, J. Astrophys. Astron., 43, 53 (2022)
\bibitem{Cap1} S. Capozziello, et al, Phys. Scr. 56, 315 (1997)
\bibitem{Cap2} G. Bimonte, S. Capozziello, V. Man'ko, G. Marmo, Phys.Rev. D58, 104009 (1998)
\bibitem{An}D.V. Anosov, Commun. Steklov Math. Inst. 90, 1 (1967)
\bibitem{Arn} V.I. Arnold, {\it Mathematical Methods of Classical Mechanics}, Springer, Berlin (1989)
\bibitem{HE} S. W. Hawking, G. F. R. Ellis,  {\it The large scale structure of space-time}, Cambridge University Press (1973)
\bibitem{Pen} R. Penrose, Rev. Mod. Phys. 37, 215 (1965)
\bibitem{MTW} C. W. Misner, K. S. Thorne, J. A. Wheeler, {\it Gravitation}, W.H. Freeman, New York (1973)
\bibitem{GK1} V.G. Gurzadyan, A.A. Kocharyan, A \& A 492, L33 (2008)
\bibitem{GK2} V.G. Gurzadyan, A.A. Kocharyan, A \& A 493, L61 (2009)
\bibitem{spot} V.G. Gurzadyan, et al, A \& A, 566, A135 (2014)
\bibitem{S2} M. Samsonyan, et al, Eur. Phys. J. Plus, 136, 350 (2021)
\bibitem{GS7} V.G. Gurzadyan A. Stepanian, Eur. Phys. J. C, 79, 169 (2019)
\bibitem{GS8} V.G. Gurzadyan, A. Stepanian, A \& A, 653, A145 (2021)
\bibitem{W} S. Weinberg, {\it Gravitation and Cosmology}, John Wiley (1972)
\bibitem{Ligo2} B. P. Abbott, et al, Nature, 551, 85, (2017)
\bibitem{Ligo3} B. P. Abbott, et al, ApJ, 909, 218 (2021)
\bibitem{BAO} N. Schoneberg, et al,  arXiv:2209.14330








\end{thebibliography}
\end{document}